\documentclass[11pt]{article}

\usepackage[margin=1in]{geometry}
\usepackage{amsmath,amssymb,bm}
\usepackage{graphicx}
\usepackage{booktabs}
\usepackage{siunitx}
\usepackage{xcolor}
\usepackage{hyperref}
\usepackage{microtype}
\usepackage{enumitem}
\usepackage{caption}
\usepackage{subcaption}

\hypersetup{colorlinks=true,linkcolor=blue,citecolor=blue,urlcolor=blue}
\newcommand{\BiCuprate}{Bi$_2$Sr$_2$CaCu$_2$O$_{8+\delta}$}
\newcommand{\Tc}{T_{\mathrm c}}

\newcommand{\ii}{\mathrm{i}}

\newcommand{\figureorplaceholder}[2][0.93\linewidth]{%
  \IfFileExists{#2}{\includegraphics[width=#1]{#2}}{%
    \fbox{\parbox[c][0.22\textheight][c]{0.88\linewidth}{\centering
      \textbf{Figure placeholder}\\[0.7em]
      File needed: \texttt{\detokenize{#2}}}}%
  }%
}

\title{\textbf{Experimental Signatures of a Memory-Dressed Cooper-Pair Field in Antinodal ARPES}}
\author{Udomsilp Pinsook*\\[0.3em]
\small Department of Physics, Faculty of Science, Chulalongkorn University, Bangkok 10330, Thailand\\
\small *email:Udomsilp.P@Chula.ac.th}
\date{\today}

\begin{document}
\maketitle

\begin{abstract}
Antinodal angle-resolved photoemission spectroscopy (ARPES) in cuprate superconductors reveals a broad incoherent background alongside a temperature-dependent superconducting contribution. We show that both features are naturally captured by a single family of parabolic cylinder functions (PCFs). First, the broad antinodal background in \BiCuprate\ is well described by a primitive PCF branch of the form $e^{-z^2/4}D_{-1/2}(z)$. Second, a superconductivity-associated component isolated via time-resolved ARPES subtraction follows a distinct $D_{3/2}$ branch. Third, the same $D_{3/2}$ structure is recovered in equilibrium temperature differences and across doping series. These findings motivate the minimal decomposition $\rho(E,T)=\rho_{\rm BG}(E,T)+\rho_{\rm SC}(E,T)$, where the superconductivity-associated spectral weight scales as $A_{\rm SC}(T)\propto |\Delta(T)|^2$.
Dynamically, these branches originate from a Cooper-pair field with finite temporal memory: a smooth memory kernel generates a leading Gaussian temporal envelope $\exp(-t^2/2\tau_G^2)$ at short times, while algebraic many-body prefactors select the PCF branch index. If the finite-memory and algebraic incoherent dressing is formally removed, the resulting structure connects to the standard Bogoliubov/BCS spectrum. 
The main conclusion is that the incoherent antinodal continuum and the longer-lived superconductivity-associated component are distinct projections of a common memory-dressed pair field.
\end{abstract}

\section{Introduction}

The superconducting gap is conventionally treated as a static order parameter, but single-particle spectroscopy in unconventional superconductors probes an object with rich temporal structure. This distinction is especially critical near the antinode of cuprate superconductors, where spectra remain broad, the pseudogap persists above $\Tc$, and a single sharp quasiparticle pole is insufficient to describe the spectral line shape \cite{Tinkham,deGennes,Timusk,Damascelli,Norman,Valla2006}. Experiments have also emphasized that pairing-related spectral changes and the pseudogap need not be identified as a single phenomenon \cite{Kondo2011}.

A powerful way to clarify this problem is to follow what experimental subtractions isolate. Equilibrium ARPES records the total occupied single-particle response \cite{Smallwood2012,He2018,Chen2022}. Pump--probe subtraction \cite{Smallwood2012}, doping comparisons suppress stationary components \cite{He2018}, and temperature differences \cite{Chen2022} isolate the spectral weight reorganized by superconductivity. Remarkably, these distinct experimental operations \cite{Smallwood2012,He2018,Chen2022} reveal a unified, compact mathematical structure.

Using a well-known special function, called parabolic cylinder functions (PCFs) \cite{Abramowitz,NIST,Pinsook2026}, the antinodal spectrum decomposes into two reproducible ingredients:
\begin{enumerate}[label=(\roman*)]
  \item A broad, asymmetric background well described by a real PCF branch $\propto e^{-z^2/4}D_{-1/2}(z)$.
  \item A superconductivity-associated component described by a higher-order $D_{3/2}$ branch that appears consistently across time-resolved ARPES, temperature evolution, and doping scaling.
\end{enumerate}

In this work, we first establish this two-component decomposition empirically. We then demonstrate that both branches arise naturally from a Cooper-pair field possessing finite temporal memory. A nonlocal temporal memory kernel $W(t-t')$ gives rise to a Gaussian cutoff $\exp(-t^2/2\tau_G^2)$, while many-body threshold dynamics fixes the PCF branch index. We further show that formal removal of the finite-memory and algebraic incoherent dressing yields the standard Bogoliubov/BCS spectral structure. This is used as a reference consistency limit, not as a claim that the observed antinodal continuum must be able to reach that state within the same material or momentum sector. In a strongly correlated antinodal regime, the physical trajectory may remain dynamically restricted to the incoherent sector.

\section{Experimental Evidence for Two Antinodal Spectral Components}
\label{sec:exp}

Throughout this work, we represent experimental spectra using the real PCF line shape
\begin{equation}
\mathcal{P}_{\nu}(E;E_0,B)
=\exp\!\left[-B^2(E_0-E)^2\right]
D_{\nu}\!\left[2B(E_0-E)\right],
\label{eq:pcfdef}
\end{equation}
with the dimensionless energy coordinate
\begin{equation}
\xi=2B(E_0-E),
\label{eq:xi}
\end{equation}
where $B$ has units of inverse energy. In the Gaussian-memory framework derived in Sec.~\ref{sec:theory}, $B$ is directly related to the effective memory time $\tau_G$ by
\begin{equation}
\tau_G=2\hbar B.
\label{eq:tauB}
\end{equation}

\subsection{Equilibrium ARPES: Robust $D_{-1/2}$ Background}

\begin{figure}[t]
\centering
\figureorplaceholder{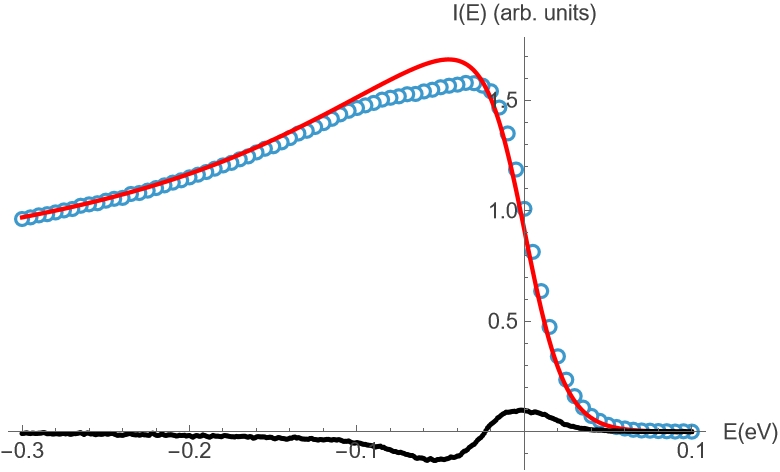}
\caption{\textbf{Equilibrium antinodal background.} Experimental antinodal ARPES spectrum of \BiCuprate\ (blue symbols) \cite{Chen2022} compared with the $D_{-1/2}$ PCF representation of Eq.~\eqref{eq:bg} (red curve). The black curve shows the residual.}
\label{fig:bg}
\end{figure}

The broad antinodal background in \BiCuprate\ can be captured by the primitive PCF branch
\begin{equation}
\rho_{\rm BG}(E)
=A_{\rm BG}\,\mathcal{P}_{-1/2}(E;E_{\rm BG},B_{\rm BG}).
\label{eq:bg}
\end{equation}
As shown in Fig.~\ref{fig:bg}, this function describes the asymmetric incoherent continuum. In temperature-dependent studies, representative parameters are $E_{\rm BG}=-0.0934$~eV and $B_{\rm BG}=4.64$~eV$^{-1}$, corresponding to an effective memory time $\tau_{\rm BG}\approx 6$~fs \cite{Pinsook2026}. This broad background persists across $\Tc$, providing an underlying continuum for the sharper superconducting features. Eq.~\eqref{eq:bg} can capture most of the spectral features with small residual (black line). A more quantitative fitting was provided in a previous work \cite{Pinsook2026}.

The significance of this representation is not merely the quality of a single fit. The $D_{-1/2}$ branch provides a compact description of the broad spectral weight that remains visible even when the sharper low-temperature superconducting structure changes substantially. In this sense, it defines a comparatively robust antinodal background against which the temperature-dependent contribution can be identified. Its persistence across $\Tc$ is also consistent with the well-known absence of a simple recovery of a conventional sharp normal-state quasiparticle at the antinode.

Within the present framework, the broadness of this component has a direct dynamical interpretation. The relatively small value of $B_{\rm BG}$, or equivalently the short effective memory time $\tau_{\rm BG}$, corresponds to a rapidly dephasing contribution in the time domain. We therefore treat the $D_{-1/2}$ sector primarily as an incoherent spectral continuum rather than as a broadened quasiparticle pole. This distinction becomes important below, because the superconductivity-associated contribution isolated independently from spectral differences belongs to a different PCF branch and possesses a substantially longer characteristic memory scale.

\subsection{Time-Resolved ARPES: Superconductivity-Associated $D_{3/2}$ Component}

\begin{figure}[t]
\centering
\figureorplaceholder{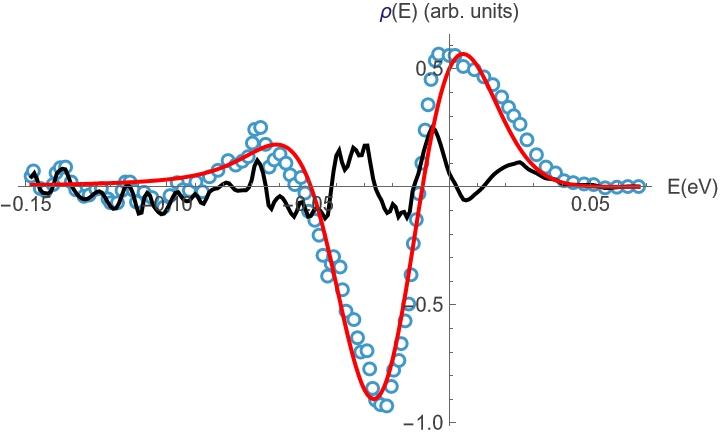}
\caption{\textbf{Superconductivity-associated component isolated by time-resolved ARPES.} Transient difference spectrum (blue symbols) \cite{Smallwood2012} fitted to the $D_{3/2}$ PCF form of Eq.~\eqref{eq:d32} (red curve). The black curve shows the residual.}
\label{fig:trarpes}
\end{figure}

In time-resolved ARPES \cite{Smallwood2012}, subtracting a pre-excitation reference spectrum removes the stationary background:
\begin{equation}
\delta I(E,t)=I(E,t)-I(E,t_0).
\label{eq:trdiff}
\end{equation}
As shown in Fig.~\ref{fig:trarpes}, the transient difference spectrum is non-monotonic and is quantitatively described by a $D_{3/2}$ PCF branch:
\begin{equation}
A_{\rm SC}
\exp[-B_{\rm SC}^2 z_-^2]D_{3/2}(2B_{\rm SC}z_-),
\label{eq:d32}
\end{equation}
where
\begin{equation}
z_-=E_{\rm SC}-E.
\label{eq:zpm}
\end{equation}
In the previous work, Smallwood et al. suggested that Eq.~\eqref{eq:trdiff} reflects a partial suppression of superconducting pairing as the pump transiently drives the system toward a more high-temperature-like electronic state. More generally, ultrafast excitation probes relaxation and energy flow in a nonequilibrium correlated state rather than a literal equilibrium-temperature trajectory \cite{VolkovKogan,Rameau2016}. Equation~\eqref{eq:d32} should therefore be interpreted as a superconductivity-associated \emph{spectral component}, rather than as the pair field itself. With the sign convention adopted below (see Appendix~\ref{app:pcf} also), the corresponding equilibrium superconducting contribution is represented with the opposite orientation in the subtraction spectrum.

The subtraction is particularly useful because spectral weight that is nearly stationary on the experimental time scale is strongly suppressed, whereas the pump-sensitive contribution is enhanced. Thus, although the transient signal is not itself a direct measurement of the pair-field propagator, it provides an experimentally controlled way of isolating a spectral component closely associated with superconducting correlations. The resulting structure is considerably more localized and sign-changing in energy than the broad $D_{-1/2}$ background, making the distinction between the two PCF sectors visible without requiring a full decomposition of the equilibrium spectrum.

Of particular importance is the appearance of the $D_{3/2}$ functional form itself. We do not infer this branch solely from temperature-dependent equilibrium fitting; it is independently exposed by a nonequilibrium subtraction procedure. The agreement therefore suggests that the $D_{3/2}$ contribution is not merely an artifact introduced by fitting the total ARPES intensity with multiple components. Rather, it represents spectral weight that responds selectively when superconducting pairing is transiently perturbed. This motivates the terminology used throughout this work: $D_{3/2}$ denotes a superconductivity-associated spectral component, while remaining distinct from the microscopic order parameter $\Delta$ and from the pair propagator $D_\Phi$ itself.

\subsection{Temperature Subtraction and Doping Evolution}

\begin{figure}[t]
\centering
\figureorplaceholder{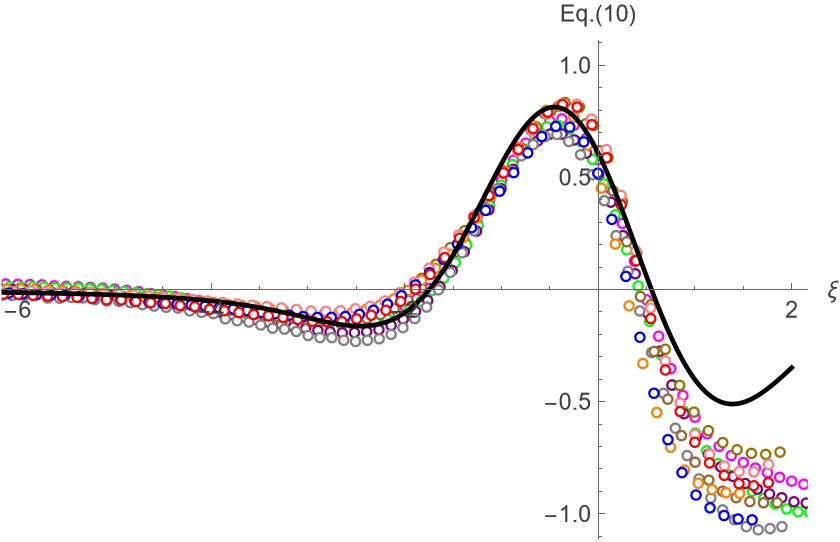}
\caption{\textbf{Doping collapse of the superconducting component.} Scaled superconductivity-induced antinodal ARPES spectra across $p=0.141$--$0.215$ (data from He \textit{et al.}~\cite{He2018}) collapsed onto the universal $D_{3/2}$ backbone (black curve) via Eq.~\eqref{eq:collapse}.}
\label{fig:tempdoping}
\end{figure}

The same $D_{3/2}$ component emerges in equilibrium temperature difference spectra:
\begin{equation}
\delta\rho_T(E)=\rho(E,T_{\rm low})-\rho(E,T_{\rm high}).
\label{eq:tempdiff}
\end{equation}
Temperature subtraction provides an equilibrium counterpart to the pump--probe analysis. When $T_{\rm low}<\Tc$ and $T_{\rm high}$ is chosen so that the superconductivity-associated contribution is substantially reduced, the slowly varying background is partly cancelled and the spectral redistribution associated with the superconducting state becomes more visible. Although temperature subtraction and ultrafast excitation are physically different operations, both suppress comparatively stationary spectral components and emphasize the part of the spectrum that changes together with superconductivity. The recovery of a similar $D_{3/2}$ structure from these two independent procedures therefore provides an important consistency check.

Furthermore, analyzing doping-dependent ARPES data across $p=0.141$--$0.215$ from He \textit{et al.} \cite{He2018} reveals a common scaling form. Changing the hole concentration modifies the characteristic energy, width, and amplitude of the superconducting spectral response, so the unscaled spectra need not coincide point by point. If these differences primarily renormalize a common underlying line shape, however, they should be largely removable by appropriate amplitude and energy rescaling. This motivates writing the isolated superconducting component as
\begin{equation}
\rho_{\rm SC}(E;p)= -A_{\rm SC}(p)\Big\{
e^{-B_{\rm SC}^2(p)z_-^2}D_{3/2}[2B_{\rm SC}(p)z_-]
+e^{-B_{\rm SC}^2(p)z_+^2}D_{3/2}[2B_{\rm SC}(p)z_+]
\Big\},
\label{eq:dopingfit}
\end{equation}
where $z_-=E_{\rm SC}(T)-E$ and $z_+=E_{\rm SC}(T)+E$.  The overall minus sign in Eq.~\eqref{eq:dopingfit} is not an independent fitting freedom. With $A_{\rm SC}\ge 0$ defined as a positive spectral-weight scale, it follows from the intrinsic orientation of the $D_{3/2}$ branch, for which $D_{3/2}(0)<0$ (Appendix~\ref{app:pcf}). Thus, the sign does not imply a negative physical density of states. 

After obtaining $A_{\rm SC}$, $B_{\rm SC}$ and $E_{\rm SC}$ via fitting, we perform function scaling by using $A_{\rm SC}(p)$ and $\xi=2B_{\rm SC}(p)[E_{\rm SC}(p)-E]$. Consequently, the occupied-side spectra collapse onto the universal backbone,
\begin{equation}
\frac{\rho_{\rm SC}(\xi;p)}{A_{\rm SC}(p)} \simeq -e^{-\xi^2/4}D_{3/2}(\xi).
\label{eq:collapse}
\end{equation}
Fig.~\ref{fig:tempdoping} shows the occupied-side spectra only. The collapse is therefore more informative than agreement with any individual unscaled spectrum: after the doping-dependent amplitude and energy scales are removed, the occupied-side data approach the same dimensionless $D_{3/2}$ backbone. The remaining discrepancies, particularly toward the opposite-energy side of the structure, are naturally expected because the complete expression in Eq.~\eqref{eq:dopingfit} contains both $+E_{\rm SC}$ and $-E_{\rm SC}$ branches, whereas only the occupied-side contribution is displayed in Fig.~\ref{fig:tempdoping}. We therefore regard the collapse as evidence for a common spectral architecture rather than exact identity of the complete spectra at different dopings. 
 
\subsection{Minimal Spectral Decomposition}
\label{sec:minimal}

\begin{figure}[htbp]
\centering
\begin{minipage}{0.48\linewidth}
\centering
\figureorplaceholder[\linewidth]{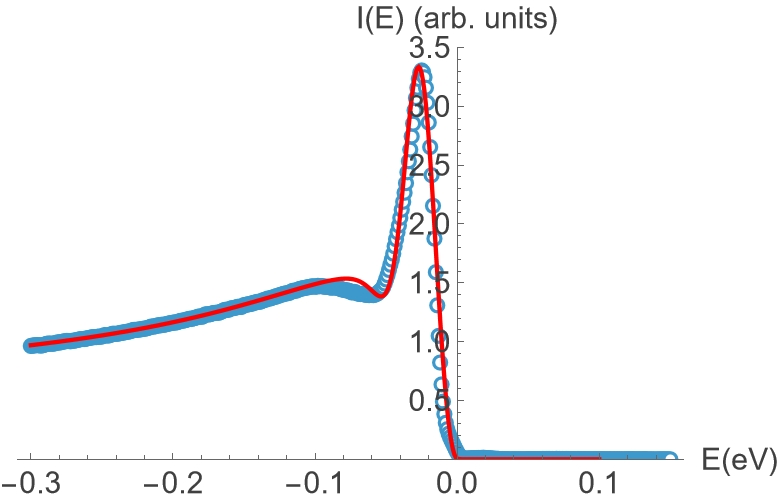}
\end{minipage}\hfill
\begin{minipage}{0.48\linewidth}
\centering
\figureorplaceholder[\linewidth]{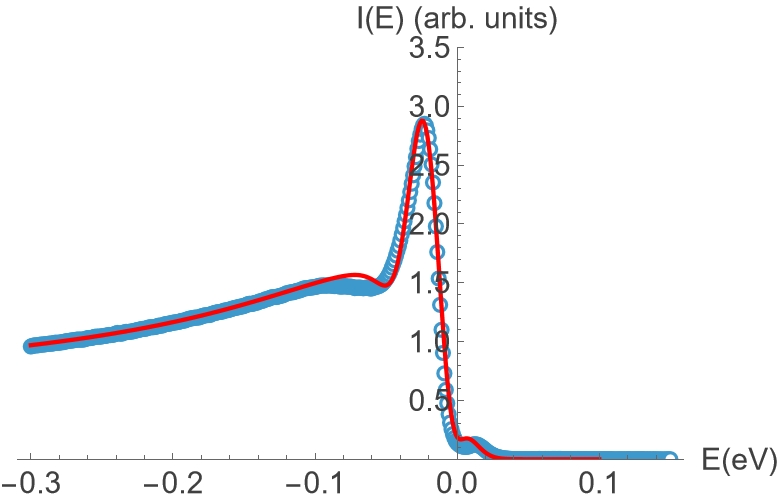}
\end{minipage}\\[0.8em]
\begin{minipage}{0.48\linewidth}
\centering
\figureorplaceholder[\linewidth]{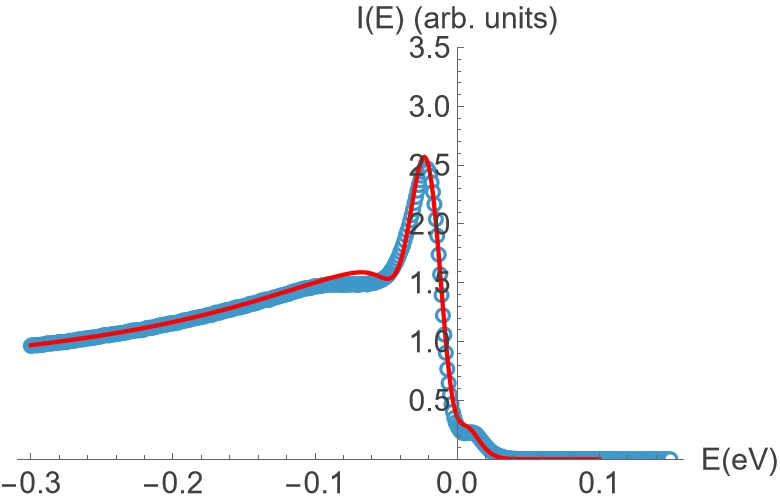}
\end{minipage}\hfill
\begin{minipage}{0.48\linewidth}
\centering
\figureorplaceholder[\linewidth]{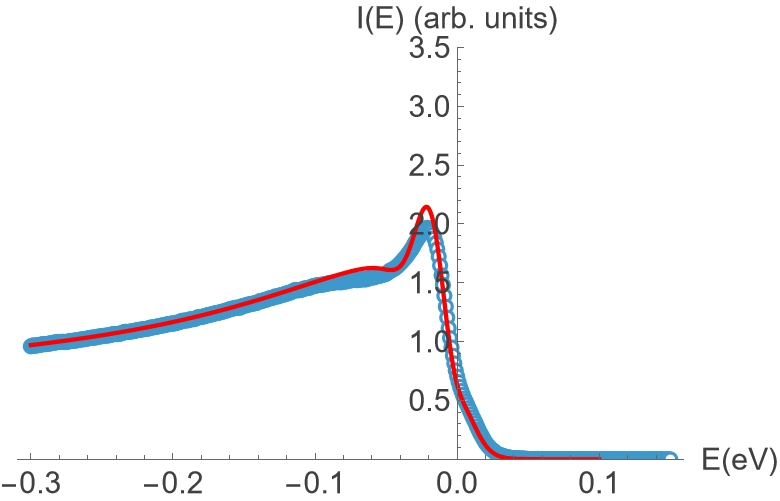}
\end{minipage}
\caption{\textbf{Temperature-dependent antinodal ARPES spectra.} Spectra at 6~K (top left), 60~K (top right), 70~K (bottom left), and 80~K (bottom right) from Chen \textit{et al.}~\cite{Chen2022} (blue points) fitted to the two-component model of Eq.~\eqref{eq:arpesintensity} (red curves) with $A_{\rm BG}=1.27$ and $A_{\rm SC}(0)=1.93$ (arb.\ units).}
\label{fig:ARPES_series}
\end{figure}

From independent observations \cite{Smallwood2012,He2018,Chen2022}, we can establish an economical two-component model for the ARPES intensity. These empirical formulas contain only six fitting parameters, written as
\begin{equation}
\rho(E,T)=\rho_{\rm BG}(E,T)+\rho_{\rm SC}(E,T),
\label{eq:minimal}
\end{equation}
where $\rho_{\rm BG}$ is the $D_{-1/2}$ background [Eq.~\eqref{eq:bg}] and $\rho_{\rm SC}$ is the $D_{3/2}$ superconductivity-associated spectral component [Eq.~\eqref{eq:dopingfit}]. Here and below, the subscript ${\rm SC}$ denotes association with superconductivity; it does not imply a conventional coherent Bogoliubov quasiparticle contribution. These $\rho$ functions are empirical energy-domain spectral components and should not be identified directly with the order parameter $\Delta$. The full ARPES intensity is modeled by
\begin{equation}
I_{\rm ARPES}(E,T) = f(E,T)\,\left[\rho_{\rm BG}(E,T)+\rho_{\rm SC}(E,T)\right]\otimes R_E,
\label{eq:arpesintensity}
\end{equation}
where $f(E,T)$ is the Fermi--Dirac distribution. $R_E$ represents instrumental energy resolution, which we set to unity in this work.

The temperature dependence is governed by the superconducting scale:
\begin{align}
E_{\rm SC}(T) &= E_{\rm SC}(0)\tanh\!\left[1.74\sqrt{\frac{T_{\rm pair}}{T}-1}\right], \label{eq:gapT}\\
A_{\rm SC}(T) &= A_{\rm SC}(0)\tanh^2\!\left[1.74\sqrt{\frac{T_{\rm pair}}{T}-1}\right], \label{eq:ampT}
\end{align}
yielding the characteristic scaling relation
\begin{equation}
A_{\rm SC}(T)\propto |\Delta(T)|^2,
\label{eq:ampdelta}
\end{equation}
while the memory parameter $B_{\rm SC} \approx 39$~eV$^{-1}$ ($\tau_{\rm SC}\approx 51$~fs) remains nearly constant over the superconducting temperature range. In addition, $E_{\rm SC}(0)=0.0213$~eV and $T_{\rm pair}= 92.7$ K.

The temperature evolution of antinodal ARPES spectra is displayed in Fig.~\ref{fig:ARPES_series}. The data demonstrate the coexistence of a robust incoherent $D_{-1/2}$ continuum with an emerging, longer-lived $D_{3/2}$ superconductivity-associated component below $\Tc$.

\section{Dynamical Formulation: Memory-Dressed Pair Field}
\label{sec:theory}
We have shown that the PCFs keep appearing in several independent experiments. In this section, we discuss a possible origin of these function forms.

\subsection{Pair Field and Nonlocal Temporal Kernel}

Applying a Hubbard--Stratonovich transformation to the Cooper channel introduces the pairing field conjugate to the fermion pair bilinear, within the standard field-theoretic formulation of superconductivity \cite{Altland}:
\begin{equation}
B_q(\tau)=\sum_k c_{-k,\downarrow}(\tau)c_{k+q,\uparrow}(\tau),
\qquad
\Delta_q(\tau)=g\sum_k\langle c_{k+q,\uparrow}(\tau)c_{-k,\downarrow}(\tau)\rangle .
\label{eq:HScompact}
\end{equation}
In Nambu space, the inverse Green function and the standard action after integrating out fermions are
\begin{equation}
\hat G^{-1}[\Delta]=
\begin{pmatrix}
\partial_\tau+\xi_k & \Delta\\
\Delta^\ast & \partial_\tau-\xi_k
\end{pmatrix},
\qquad
S_{\rm eff}[\Delta] = \int_0^\beta d\tau\sum_q\frac{|\Delta_q|^2}{g} - \mathrm{Tr}\ln \hat G^{-1}[\Delta].
\label{eq:Seff}
\end{equation}
To account for retardation from collective degrees of freedom that are not retained explicitly in the reduced pair-field description, we introduce a bilocal temporal memory action:
\begin{equation}
S_{\rm mem}[\Delta] = -\frac{1}{2}\int d1\,d2\, |\Delta(1)|^2\,W(1-2)\,|\Delta(2)|^2,
\label{eq:Smem}
\end{equation}
where $W(t-t')$ is a stationary, even memory kernel regular at short times. Its dependence only on the time difference expresses time-translation invariance: the absolute origin of time has no significance in the stationary formulation. Such a kernel can arise by coarse-graining an unresolved collective sector of the many-body system; a minimal microscopic construction is given in Appendix~\ref{app:kernelorigin}. No specific microscopic mode is required for the results below.

\subsection{Emergence of the Gaussian Temporal Envelope}

We distinguish the pairing field $\Delta$ from its dynamical fluctuations. We denote the latter by $\Phi$ and their causal propagator by $D_\Phi$; the symbol $G_e$ is reserved for the electronic Green function measured indirectly by ARPES. Throughout the following stationary formulation, the argument $t$ in $D_\Phi(t)$ denotes the \emph{relative correlation time} between two events, after choosing one event as the time origin. It should therefore not be interpreted as an absolute laboratory time or, in a pump--probe experiment, as the pump--probe delay. In cumulant expansion, the memory-dressed pair-field propagator is expressed as
\begin{equation}
D_{\Phi}(t)=D_{\rm MB}(t)\,e^{-\mathcal F(t)},
\label{eq:cumulantD}
\end{equation}
with the accumulated memory functional
\begin{equation}
\mathcal F(t) = \int_0^t ds\int_0^s ds'\,\mathcal W(s-s') = \int_0^t d\tau\,(t-\tau)\,\mathcal W(\tau).
\label{eq:FfromW}
\end{equation}
Expanding the regular kernel $\mathcal W(\tau)=\mathcal W(0)+\mathcal O(\tau^2)$ yields to leading order
\begin{equation}
\mathcal F(t) = \frac{\mathcal W(0)}{2}t^2 + \mathcal O(t^4) \equiv \frac{t^2}{2\tau_G^2},
\label{eq:Fshort}
\end{equation}
where $\tau_G^{-2}\equiv \mathcal W(0)$. Accordingly, $\tau_G$ characterizes the decay of temporal correlations in relative time; it should not be identified directly with an absolute experimental clock time or a pump--probe relaxation time. Hence, the Gaussian envelope
\begin{equation}
M(t) \equiv e^{-\mathcal F(t)} \simeq \exp\!\left[-\frac{t^2}{2\tau_G^2}\right]
\label{eq:GaussianFromW}
\end{equation}
is the leading short-time consequence of a smooth bilocal memory kernel. Representing the bare many-body propagation as $D_{\rm MB}(t)=D_0(\ii t/\tau_0)^{-a}e^{-\ii\Omega_0 t}$, the full memory-dressed correlator becomes
\begin{equation}
D_{\Phi}(t)=D_0\left(\frac{\ii t}{\tau_0}\right)^{-a} e^{-\ii\Omega_0 t}\exp\!\left[-\frac{t^2}{2\tau_G^2}\right].
\label{eq:masterD}
\end{equation}

The energy-domain pair-field response follows from the one-sided causal Fourier transform of this correlator. Here the restriction to $t>0$ expresses the retarded response convention rather than an occupied-energy restriction; positive- and negative-energy spectral structures, when both are required, arise from the corresponding frequency components of the causal response. The transform of Eq.~\eqref{eq:masterD} requires evaluating
\begin{equation}
\int_0^\infty dt\,t^{-a}\exp\!\left[-\frac{t^2}{2\tau_G^2}+\ii(\omega-\Omega_0)t\right],
\label{eq:PCFintegral}
\end{equation}
which evaluates in closed form to parabolic cylinder functions (Appendix~\ref{app:pcf}). The background branch corresponds to $a=1/2$ ($D_{-1/2}$), representing the incoherent continuum, while $a=-3/2$ produces the $D_{3/2}$ resonant pairing component.

\subsection{Projection onto the ARPES Single-Particle Spectrum}
\label{sec:pairtoarpes}

ARPES probes the single-particle spectral function $A_e(\bm k,\omega) = -\frac{1}{\pi}\operatorname{Im}G_e^R(\bm k,\omega)$. The pair field contributes to the \emph{electronic} self-energy through the convolution
\begin{equation}
\Sigma_e^{(\Phi)}(\bm k,\omega) = \sum_{\bm q}\int\frac{d\Omega}{2\pi}\, D_\Phi(\bm q,\Omega)\, G_h(\bm k-\bm q,\omega-\Omega),
\label{eq:pairsigma-general}
\end{equation}
where the superscript $(\Phi)$ emphasizes that this is the contribution to the electron self-energy generated by the pair field, rather than a self-energy of the pair propagator itself. This establishes the dynamical bridge $D_\Phi \to \Sigma_e^{(\Phi)} \to G_e \to A_e$. The antinodal region is particularly useful spectroscopically because coherent single-particle quasiparticle poles are strongly suppressed, especially in the pseudogap regime. It therefore provides a comparatively quiet momentum-space window in which the broad self-energy contribution generated by the pair field can be exposed with less competition from coherent quasiparticle spectral weight. A weak underlying dispersion may further enhance this visibility, but a flat-band condition is not required for the emergence of the PCF line shape. As discussed in Sec.~\ref{sec:bcslimit}, a formal removal of the incoherent dressing connects this construction to the conventional BCS form, although that coherent reference limit need not be physically accessible from the cuprate antinodal regime.

\subsection{Physical Interpretation and Relation to a Coherent Bogoliubov/BCS Reference Limit}
\label{sec:bcslimit}

At this stage, we summarize the characteristic real-time dynamics associated with the two empirical spectral components. To keep the order parameter distinct from its dynamical correlator, we reserve $\Delta$ for the Hubbard--Stratonovich/BCS pairing field and denote the real causal projection of the memory-dressed pair propagator by $\mathcal P_\Phi(t,T)\equiv\operatorname{Re}D_\Phi(t,T)$. A compact phenomenological representation is
\begin{align}
\mathcal P_{\Phi}(t,T) = &\,C_{\rm BG}(T)t^{-1/2}\exp\!\left[-\frac{t^2}{2\tau_{\rm BG}^2}\right]\cos(\Omega_{\rm BG}t+\phi_{\rm BG}) \nonumber\\
&+ C_{\rm SC}(T)t^{3/2}\exp\!\left[-\frac{t^2}{2\tau_{\rm SC}^2}\right]\cos(\Omega_{\rm SC}t+\phi_{\rm SC}).
\label{eq:twotime}
\end{align}
The short memory time $\tau_{\rm BG}\approx 6$~fs reflects the rapidly dephasing pseudogap continuum, whereas $\tau_{\rm SC}\approx 51$~fs reflects the longer-lived pairing correlations. Because the Fourier-transform normalization of the $D_{3/2}$ sector depends on its algebraic index and memory scale, and $\tau_{\rm SC}$ is nearly temperature independent over the fitted range, the time-domain amplitude carries the same temperature dependence up to an approximately temperature-independent normalization factor, $C_{\rm SC}(T)\propto A_{\rm SC}(T)\propto |\Delta(T)|^2$. Equation~\eqref{eq:twotime} is intended to summarize the characteristic algebraic powers, oscillation scales, and finite-memory envelopes associated with the $D_{-1/2}$ and $D_{3/2}$ sectors. It should not be interpreted as a unique inverse transform of the full occupied- and unoccupied-side decomposition in Eq.~\eqref{eq:dopingfit}; in particular, the latter contains both $+E_{\rm SC}$ and $-E_{\rm SC}$ spectral branches explicitly. Thus Eq.~\eqref{eq:twotime} is the real-time phenomenological counterpart of the two-component decomposition in Eq.~\eqref{eq:minimal}, rather than a term-by-term inversion of Eq.~\eqref{eq:dopingfit}.

To establish a formal connection with conventional superconductivity, we now consider a separate idealized reference regime in which the finite-memory and algebraic incoherent dressing is absent, consistent with the familiar coherent pairing framework and its collective/instability formulation \cite{Altland,Anderson,Thouless}. This construction should not be interpreted as a physically required endpoint of the antinodal $D_{3/2}$ continuum. In particular, the correlated antinodal state may occupy a dynamical sector from which the conventional coherent limit is not experimentally accessible. We do not derive such an obstruction here; rather, we avoid assuming that a continuous path between the two regimes must exist. In the idealized coherent reference case, the pair field is represented by the static superconducting order parameter $\Delta_{\bm k}$, and the electronic Nambu Green function takes the conventional BCS form:
\begin{equation}
\hat G^{-1}_0(\bm k,\omega)=
\begin{pmatrix}
\omega-\xi_{\bm k}+\ii0^+ & -\Delta_{\bm k}\\
-\Delta_{\bm k}^\ast & \omega+\xi_{\bm k}+\ii0^+
\end{pmatrix}.
\label{eq:BCSnambu}
\end{equation}
The normal component is
\begin{equation}
G_{11}^R(\bm k,\omega) = \frac{\omega+\xi_{\bm k}}{(\omega+\ii0^+)^2-\xi_{\bm k}^2-|\Delta_{\bm k}|^2} = \frac{u_{\bm k}^2}{\omega-E_{\bm k}+\ii0^+} + \frac{v_{\bm k}^2}{\omega+E_{\bm k}+\ii0^+},
\label{eq:G11poles}
\end{equation}
where $E_{\bm k}=\sqrt{\xi_{\bm k}^2+|\Delta_{\bm k}|^2}$, and $u_{\bm k}^2, v_{\bm k}^2 = \frac{1}{2}(1\pm \xi_{\bm k}/E_{\bm k})$. The spectral function reduces to sharp quasiparticle poles:
\begin{equation}
A(\bm k,E) = u_{\bm k}^2\delta(E-E_{\bm k}) + v_{\bm k}^2\delta(E+E_{\bm k}).
\label{eq:bcsspectral}
\end{equation}
In this idealized coherent case, the pair correlator becomes $D_\Phi^{\rm coh}(\bm q,\Omega)\propto |\Delta_{\bm k}|^2\delta_{\bm q,0}2\pi\delta(\Omega)$, reducing Eq.~\eqref{eq:pairsigma-general} to the standard BCS electron self-energy $\Sigma_{e,\rm BCS}^{(\Phi)}(\bm k,\omega)=|\Delta_{\bm k}|^2/(\omega+\xi_{\bm k}+\ii0^+)$. The corresponding spectral function contains both the occupied and unoccupied Bogoliubov branches in Eq.~\eqref{eq:bcsspectral}. The purpose of this construction is therefore to demonstrate algebraic compatibility with the familiar coherent Bogoliubov structure, rather than to prescribe a microscopic crossover from the measured antinodal PCF continuum to that state. A formal limiting solution and a physically accessible material trajectory are distinct statements; the present work establishes only the former.

\subsection{Tunneling Spectroscopy: Density-of-States Cross-Check}

\begin{figure}[htbp]
\centering
\figureorplaceholder[0.8\linewidth]{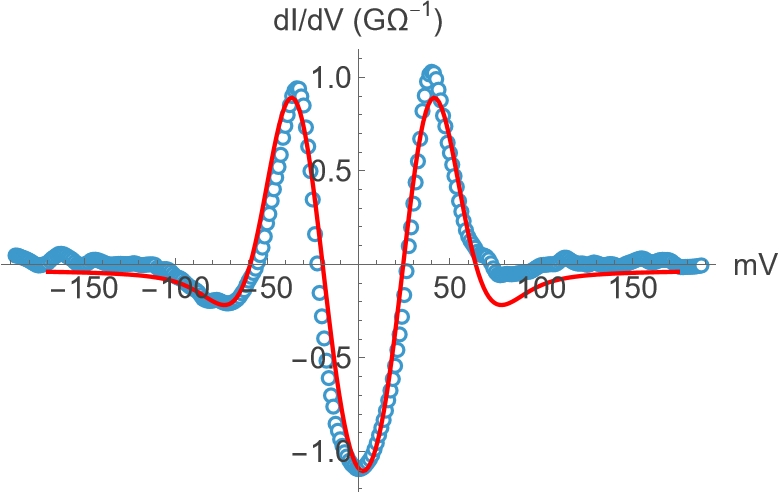}
\caption{\textbf{Differential tunneling conductance.} Background-subtracted $dI/dV$ data in \BiCuprate\ from Renner and Fischer~\cite{Renner1995} (blue points) compared with the $D_{3/2}$ PCF model of Eq.~\eqref{eq:dopingfit} using the simple scaling of Eq.~\eqref{eq:tunneling_fit} (red curve).}
\label{fig:Tunneling}
\end{figure}

Scanning tunneling spectroscopy (STS) probes a momentum-integrated local density of states (LDOS), and therefore provides a useful cross-probe test of whether the same empirical spectral structure identified in ARPES can also describe a local spectroscopic response \cite{Renner1995}. In the standard low-temperature tunneling approximation,
\begin{equation}
\frac{dI}{dV}(V) \propto N(eV) = \sum_{\mathbf{k}} A(\mathbf{k}, eV),
\end{equation}
where $V$ is the sample bias voltage. Because STS integrates over momentum, agreement with the same $D_{3/2}$ line shape should be interpreted as a consistency test rather than as a direct momentum-space identification of the antinodal contribution.

To test this connection, we model the background-subtracted differential conductance using the symmetric $D_{3/2}$ spectral component defined in Eq.~\eqref{eq:dopingfit}, subject to a constant offset and an energy scaling factor:
\begin{equation}
\frac{dI}{dV}(V) \simeq \alpha_0 + \alpha_1 \rho_{\mathrm{SC}}(s \cdot eV + E_{\mathrm{off}}),
\label{eq:tunneling_fit}
\end{equation}
where $\alpha_0$ and $\alpha_1$ set the baseline and spectral amplitude, $s$ is an effective dimensionless energy scaling factor, and $E_{\mathrm{off}}$ accounts for a small energy offset or particle--hole asymmetry \cite{Renner1995}.

As shown in Fig.~\ref{fig:Tunneling}, the low-temperature differential conductance of $\mathrm{Bi_{2}Sr_{2}CaCu_{2}O_{8+\delta}}$ from Renner and Fischer \cite{Renner1995} is reproduced well by the scaled $D_{3/2}$ profile over the fitted bias window, including the coherence-peak and dip structure. We regard this agreement as a nontrivial cross-probe consistency check: the same PCF branch used for the superconductivity-associated ARPES component also captures characteristic structure in a momentum-integrated tunneling spectrum. It does not by itself establish that the complete tunneling DOS is generated solely by the antinodal pair-field sector, nor does it independently determine the microscopic origin of the memory kernel.

\section{Discussion}
\label{sec:discussion}

The central empirical result of this work is the recurrence of two PCF structures in antinodal spectroscopy: a broad $D_{-1/2}$ background and a more structured $D_{3/2}$ superconductivity-associated component. The antinode is especially useful as a spectroscopic window because coherent single-particle poles are strongly suppressed there in the pseudogap regime, reducing their competition with the broad self-energy response. This interpretation is consistent with the long-established loss of a conventional antinodal Fermi-surface pole and with experiments separating pseudogap and pairing-related spectral evolution \cite{Norman,Valla2006,Kondo2011}. The ``quiet'' antinodal window is therefore a condition for visibility rather than the microscopic origin of the PCF structure; in particular, a flat-band condition is not required.

The memory-dressed formulation should also be distinguished from a broadened BCS ansatz. The finite temporal envelope and algebraic prefactor define an incoherent dynamical regime whose characteristic line shapes are the PCF branches. Conventional Bogoliubov theory, by contrast, describes a coherent reference regime with sharp particle--hole poles. Both descriptions are rooted in Cooper pairing, but the present work does not derive a unique continuous trajectory connecting them. This situation is familiar in many-body theory, where distinct asymptotic formulations can be valid in different dynamical regimes even when they describe the same underlying degrees of freedom. Our coherent construction in Sec.~\ref{sec:bcslimit} therefore establishes compatibility with the conventional pairing framework \cite{Altland,Anderson,Thouless}, not an absolute reduction of the measured cuprate antinodal state to BCS.

The unresolved theoretical problem is the bridge between these regimes. A complete theory would have to determine self-consistently how the memory kernel, algebraic many-body dressing, and phase coherence evolve together, and whether a coherent Bogoliubov sector is physically reachable at a given momentum and in a given material. The bridge could be smooth, strongly momentum selective, material dependent, or dynamically obstructed; the present data do not distinguish these possibilities. Nonequilibrium relaxation studies illustrate why this distinction matters: transient suppression of pairing need not be equivalent to heating the system through a sequence of equilibrium states \cite{Smallwood2012,VolkovKogan,Rameau2016}.

Finally, the bilocal kernel is intentionally coarse grained. Internal charge, spin, lattice, plasmonic, or particle--hole degrees of freedom can all contribute to the unresolved collective sector, and charge-order correlations are experimentally established in the cuprates \cite{Miao2021}. Identifying which microscopic channels dominate $W(t-t')$, and whether the same channels control the crossover toward or away from coherent quasiparticles, remains an open problem. The present work is therefore best viewed as establishing the finite-memory spectroscopic regime and its experimentally testable structure, while leaving the dynamical bridge to conventional coherence as a separate theoretical task.

\section{Conclusion}

Antinodal spectroscopy in $\mathrm{Bi_{2}Sr_{2}CaCu_{2}O_{8+\delta}}$ separates into a robust $D_{-1/2}$ incoherent continuum and a superconductivity-associated $D_{3/2}$ component across equilibrium ARPES, transient pump--probe subtraction, temperature evolution, doping scaling, and a cross-probe tunneling comparison \cite{Smallwood2012,He2018,Chen2022,Renner1995}. Both PCF branches are naturally generated within a Cooper-pair description possessing finite temporal memory: a smooth bilocal kernel produces the leading short-time Gaussian envelope $\exp(-t^2/2\tau_G^2)$, while algebraic many-body dynamics selects the PCF branch index. The extracted scales distinguish a rapidly dephasing background ($\tau_{\rm BG}\approx6$~fs) from a longer-lived superconductivity-associated response ($\tau_{\rm SC}\approx51$~fs), with $A_{\rm SC}(T)\propto|\Delta(T)|^2$ over the fitted temperature range.

When the finite-memory and algebraic incoherent dressing is removed by construction, the formulation is compatible with the familiar coherent Bogoliubov/BCS spectral structure. This reference limit should not be confused with a derived physical crossover of the cuprate antinode. The present work establishes the finite-memory regime and its spectroscopic signatures; a complete theory would additionally require the dynamical bridge, if physically accessible, between this correlated incoherent sector and conventional phase-coherent Bogoliubov quasiparticles. Determining whether that bridge is smooth, momentum selective, material dependent, or dynamically inaccessible is left as an open problem. More broadly, the same finite-memory scaling structure may provide a route toward understanding why related PCF line shapes and marginal spectral collapses can recur across distinct material classes \cite{MarginalCollapse}. Such transferability should be regarded as an empirical indication of a shared dynamical regime, rather than as evidence that the underlying microscopic mechanisms are identical in different materials.

\appendix
\section{Fourier Transform and Real-Quadrature PCF Branch Selection}
\label{app:pcf}

In this Appendix, we show how the one-sided causal Fourier transform of the Gaussian-damped power law leads to the real parabolic cylinder fitting functions used throughout the text, and how an appropriate real-quadrature projection isolates the fitting branch used in the analysis. The phase introduced below is a mathematical projection convention for the complex causal representation and should not be identified with a physically measured superconducting gauge phase.

\subsection{Complex Causal Fourier Transform}

Consider the memory-dressed pair-field propagator with an auxiliary projection phase $\phi$ used to select a real quadrature of the complex causal response. As in the main text, $t$ denotes relative correlation time rather than absolute laboratory time:
\begin{equation}
D_\Phi(t;\phi) = D_0 \left(\frac{\ii t}{\tau_0}\right)^p \exp\!\left[-\frac{t^2}{2\tau_G^2}\right] e^{-\ii\Omega_0 t}\, e^{\ii\phi}, \qquad p > -1.
\label{eq:D_with_phase}
\end{equation}
For the retarded response, causality is implemented by $D_\Phi^R(t)=\theta(t)D_\Phi(t)$, so the Fourier transform is naturally one-sided,
\begin{equation}
F_p(q;\phi) = \int_0^\infty dt\, D_\Phi(t;\phi)\, e^{\ii\omega t}, \qquad q \equiv \omega - \Omega_0.
\end{equation}
The negative-time interval is therefore not a missing identical copy of the response, and no additional factor of two is implied. This causal restriction should also not be identified with an occupied-side restriction: the $+E$ and $-E$ spectral branches originate from positive- and negative-frequency components within the causal response itself. Substituting the dimensionless variables $u = t/\tau_G$ and $x = q\tau_G$, we obtain:
\begin{equation}
F_p(q;\phi) = C_0\, e^{\ii\phi}\,\tau_G^{p+1} \int_0^\infty du\, u^p \exp\!\left(-\frac{u^2}{2}\right) e^{\ii xu}.
\end{equation}
Using the standard integral identity for the parabolic cylinder function \cite{Abramowitz,NIST}, the integral evaluates to
\begin{equation}
F_p(q;\phi) = \mathcal{N}_p\, e^{\ii\phi} \exp\!\left(-\frac{x^2}{4}\right) D_{-(p+1)}(-\ii x),
\label{eq:Fp_imag}
\end{equation}
where $\mathcal{N}_p = C_0\,\Gamma(p+1)\,\tau_G^{p+1}$. Note that because $x = q\tau_G$ is real, the argument of the parabolic cylinder function in Eq.~\eqref{eq:Fp_imag} is purely imaginary ($-\ii x$).

\subsection{Transformation from Imaginary Argument to Real Functions}

To express Eq.~\eqref{eq:Fp_imag} in terms of standard real Weber functions of a real variable $x$, we apply the parabolic cylinder reflection identity (NIST/DLMF 12.2.17) \cite{NIST}:
\begin{equation}
D_{-(p+1)}(-\ii x) = \frac{\Gamma(-p)}{\sqrt{2\pi}} \left[ e^{-\ii\pi(p+1)/2} D_p(x) + e^{\ii\pi(p+1)/2} D_p(-x) \right].
\label{eq:reflection}
\end{equation}
Substituting Eq.~\eqref{eq:reflection} into Eq.~\eqref{eq:Fp_imag} yields the general decomposition:
\begin{equation}
F_p(q;\phi) = \mathcal{A}_p \exp\!\left(-\frac{x^2}{4}\right) \left[ e^{\ii\left(\phi - \frac{\pi(p+1)}{2}\right)} D_p(x) + e^{\ii\left(\phi + \frac{\pi(p+1)}{2}\right)} D_p(-x) \right],
\label{eq:two_branches}
\end{equation}
where $\mathcal{A}_p = \mathcal{N}_p\,\Gamma(-p)/\sqrt{2\pi}$.

\subsection{Real-Quadrature Selection and Exact Cancellation}

To extract the real PCF quadrature used for fitting, we choose the auxiliary projection phase
\begin{equation}
\phi = \phi_+ \equiv \frac{\pi(p+1)}{2},
\end{equation}
the two phase factors in Eq.~\eqref{eq:two_branches} become:
\begin{align}
e^{\ii\left(\phi_+ - \frac{\pi(p+1)}{2}\right)} &= e^{\ii 0} = 1, \\
e^{\ii\left(\phi_+ + \frac{\pi(p+1)}{2}\right)} &= e^{\ii\pi(p+1)} = \cos[\pi(p+1)] + \ii\sin[\pi(p+1)].
\end{align}
Crucially, for the half-integer powers relevant to this work:
\begin{itemize}
  \item For $p = -1/2$ (background continuum): $\pi(p+1) = \pi/2 \implies \cos(\pi/2) = 0$.
  \item For $p = 3/2$ (superconducting peak): $\pi(p+1) = 5\pi/2 \implies \cos(5\pi/2) = 0$.
\end{itemize}
Consequently, taking the real projection identically cancels the counter-propagating $D_p(-x)$ term:
\begin{equation}
\operatorname{Re}\left[ F_p(q;\phi_+) \right] = \mathcal{A}_p \exp\!\left(-\frac{x^2}{4}\right) D_p(x).
\label{eq:real_PCF_final}
\end{equation}
Mapping the dimensionless coordinate to the real energy shift via $x = \xi = 2B(E_0 - E)$ directly produces the real fitting branches $\mathcal{P}_{-1/2}$ and $\mathcal{P}_{3/2}$ defined in Eq.~\eqref{eq:pcfdef}.

\subsection{Intrinsic Sign of the $p=3/2$ Branch}
\label{app:sign}

The relative orientation of the two half-integer branches is fixed by the parabolic cylinder function itself. At the branch center,
\begin{equation}
D_p(0)=\frac{2^{p/2}\sqrt{\pi}}{\Gamma[(1-p)/2]}.
\label{eq:Dp0}
\end{equation}
For the background branch,
\begin{equation}
D_{-1/2}(0)=\frac{2^{-1/4}\sqrt{\pi}}{\Gamma(3/4)}>0,
\end{equation}
whereas for the superconductivity-associated branch,
\begin{equation}
D_{3/2}(0)=\frac{2^{3/4}\sqrt{\pi}}{\Gamma(-1/4)}<0,
\label{eq:D32negative}
\end{equation}
because $\Gamma(-1/4)<0$. Therefore, if the amplitudes $A_{\rm BG}$ and $A_{\rm SC}$ are defined as non-negative spectral-weight scales, the naturally oriented real branches are
\begin{equation}
\mathcal{P}^{(+)}_p(\xi)
=s_p\,e^{-\xi^2/4}D_p(\xi),
\qquad
s_p\equiv \operatorname{sgn}[D_p(0)],
\label{eq:orientedPCF}
\end{equation}
so that
\begin{equation}
\mathcal{P}^{(+)}_{-1/2}(\xi)
=+e^{-\xi^2/4}D_{-1/2}(\xi),
\qquad
\mathcal{P}^{(+)}_{3/2}(\xi)
=-e^{-\xi^2/4}D_{3/2}(\xi).
\label{eq:signbranches}
\end{equation}
The minus sign used in Eqs.~\eqref{eq:dopingfit} and \eqref{eq:collapse} therefore follows from the intrinsic sign of the $p=3/2$ PCF branch once a positive amplitude convention is adopted; it is not introduced as an additional fitting parameter. This sign convention concerns the orientation of the isolated spectral component and should not be interpreted as a negative physical density of states.

\section{Microscopic Origin of the Bilocal Memory Kernel}
\label{app:kernelorigin}

Equation~\eqref{eq:Smem} is used in the main text as an effective description and does not require identification of a unique microscopic mode. For completeness, a minimal construction can be written by separating the explicitly retained fermionic Cooper sector from an unresolved collective sector,
\begin{equation}
H=H_f+H_C+H_X+H_{C-X},
\label{eq:Hminimal}
\end{equation}
with
\begin{align}
H_f &= \sum_{\bm k\sigma}\xi_{\bm k}c_{\bm k\sigma}^\dagger c_{\bm k\sigma},\\
H_C &= -g\sum_{\bm q}B_{\bm q}^\dagger B_{\bm q},
\qquad
B_{\bm q}=\sum_{\bm k}c_{-\bm k,\downarrow}c_{\bm k+\bm q,\uparrow}.
\end{align}
The variables $X_\lambda$ represent dynamical degrees of freedom that are not kept explicitly. For a Gaussian collective sector one may take, for example,
\begin{equation}
H_X=\sum_\lambda\left(\frac{P_\lambda^2}{2M_\lambda}+\frac{M_\lambda\omega_\lambda^2X_\lambda^2}{2}\right).
\label{eq:HX}
\end{equation}
At the coarse-grained pair-field level, the leading gauge-invariant coupling is to the pair intensity,
\begin{equation}
S_{C-X}=\int d\tau\sum_\lambda g_\lambda X_\lambda(\tau)|\Delta(\tau)|^2.
\label{eq:SCX}
\end{equation}
Integrating out the Gaussian variables $X_\lambda$ gives a nonlocal contribution of the form
\begin{equation}
S_{\rm mem}[\Delta]= -\frac12\int d\tau\,d\tau'\,|\Delta(\tau)|^2\,W(\tau-\tau')\,|\Delta(\tau')|^2,
\label{eq:SmemAppendix}
\end{equation}
where, schematically,
\begin{equation}
W(\tau-\tau')=\sum_{\lambda\lambda'}g_\lambda g_{\lambda'}\,\langle X_\lambda(\tau)X_{\lambda'}(\tau')\rangle.
\label{eq:WfromX}
\end{equation}
Thus the memory kernel records the temporal correlations of whatever sector has been coarse-grained out. The fields $X_\lambda$ need not constitute an external bath: in a correlated electronic material they may represent internal charge or spin fluctuations, plasmonic modes, lattice degrees of freedom, particle--hole continua, or combinations thereof. Charge-order fluctuations are one experimentally established example of an internal collective sector in cuprates \cite{Miao2021}. We deliberately leave this sector unspecified because the subsequent Gaussian-memory result requires only that the effective kernel be stationary and regular at short times, not that one particular microscopic fluctuation mechanism be selected.

\section*{Acknowledgment}
I dedicate whatever merit this work may have to Professor N. M. Temme, whose generous mathematical guidance in 2008 unexpectedly finds an echo in the present work. I remain sincerely grateful for his generosity, insight, and guidance.

\section*{AI Declaration}
During the preparation of this work, the authors used ChatGPT (OpenAI) for language editing only. The authors reviewed and edited the output as needed and take full responsibility for the content of the published article.


\begin{thebibliography}{99}

\bibitem{Tinkham}
M. Tinkham, \textit{Introduction to Superconductivity}, 2nd ed. (McGraw-Hill, New York, 1996).

\bibitem{deGennes}
P. G. de Gennes, \textit{Superconductivity of Metals and Alloys} (Westview Press, Boulder, 1999).

\bibitem{Timusk}
T. Timusk and B. Statt, The pseudogap in high-temperature superconductors: an experimental survey, Rep. Prog. Phys. \textbf{62}, 61 (1999).

\bibitem{Damascelli}
A. Damascelli, Z. Hussain, and Z.-X. Shen, Angle-resolved photoemission studies of the cuprate superconductors, Rev. Mod. Phys. \textbf{75}, 473 (2003).

\bibitem{Norman}
M. R. Norman \textit{et al.}, Destruction of the Fermi surface in underdoped high-$T_c$ superconductors, Nature \textbf{392}, 157 (1998).

\bibitem{Valla2006}
T. Valla \textit{et al.}, The ground state of the pseudogap in cuprate superconductors, Science \textbf{314}, 1914--1916 (2006).

\bibitem{Kondo2011}
T. Kondo \textit{et al.}, Disentangling Cooper-pair formation above the transition temperature from the pseudogap state in the cuprates, Nat. Phys. \textbf{7}, 21--25 (2011).

\bibitem{Smallwood2012}
C. L. Smallwood, J. P. Hinton, C. Jozwiak, W. Zhang, J. D. Koralek, H. Eisaki, D.-H. Lee, J. Orenstein, and A. Lanzara,
Tracking Cooper Pairs in a Cuprate Superconductor by Ultrafast Angle-Resolved Photoemission, Science \textbf{336}, 1137--1139 (2012).

\bibitem{He2018}
Y. He \textit{et al.}, Rapid change of superconductivity and electron-phonon coupling through critical doping in Bi-2212, Science \textbf{362}, 62 (2018).

\bibitem{Chen2022}
S.-D. Chen \textit{et al.}, Unconventional spectral signature of $T_c$ in a pure $d$-wave superconductor, Nature \textbf{601}, 562--567 (2022).

\bibitem{Abramowitz}
M. Abramowitz and I. A. Stegun, Eds., \textit{Handbook of Mathematical Functions with Formulas, Graphs, and Mathematical Tables} (Dover, New York, 1965).

\bibitem{NIST}
F. W. J. Olver, D. W. Lozier, R. F. Boisvert, and C. W. Clark, Eds., \textit{NIST Handbook of Mathematical Functions} (Cambridge University Press, New York, 2010).

\bibitem{Pinsook2026}
U. Pinsook, ARPES of Bi$_2$Sr$_2$CaCu$_2$O$_{8+\delta}$ interpreted via a particle in a system of dynamic scatterers, J. Phys.: Condens. Matter \textbf{38}, 125603 (2026).

\bibitem{VolkovKogan}
A. F. Volkov and Sh. M. Kogan, Collisionless relaxation of the energy gap in superconductors, Sov. Phys. JETP \textbf{38}, 1018 (1974).

\bibitem{Rameau2016}
J. D. Rameau \textit{et al.}, Energy dissipation from a correlated system driven out of equilibrium, Nat. Commun. \textbf{7}, 13761 (2016).

\bibitem{Altland}
A. Altland and B. Simons, \textit{Condensed Matter Field Theory}, 2nd ed. (Cambridge University Press, Cambridge, 2010).

\bibitem{Anderson}
P. W. Anderson, Random-phase approximation in the theory of superconductivity, Phys. Rev. \textbf{112}, 1900 (1958).

\bibitem{Thouless}
D. J. Thouless, Perturbation theory in statistical mechanics and the theory of superconductivity, Ann. Phys. \textbf{10}, 553 (1960).

\bibitem{Renner1995}
Ch. Renner and \O. Fischer, Vacuum tunneling spectroscopy and asymmetric density of states of Bi$_2$Sr$_2$CaCu$_2$O$_{8+\delta}$, Phys. Rev. B \textbf{51}, 9208--9218 (1995).

\bibitem{Miao2021}
H. Miao \textit{et al.}, Charge density waves in cuprate superconductors beyond the critical doping, npj Quantum Materials \textbf{6}, 31 (2021).

\bibitem{MarginalCollapse}
U. Pinsook, P. Tasee, and J. Seeyangnok, Evidence of universal spectral collapse at a marginal dynamical regime, arXiv:2603.09665 [cond-mat.str-el] (2026). 

\end{thebibliography}
\end{document}